# The Gibbs paradox in classical thermodynamics is a consequence of the erroneous attribution of the entropy of an ideal gas to additive quantities

Volodymyr Ihnatovych

Igor Sikorsky Kyiv Polytechnic Institute

e-mail: v.ihnatovych@kpi.ua

## Abstract

The article reveals the error that in classical thermodynamics leads to the Gibbs paradox. The essence of the error lies in the fact that the entropy of an ideal gas is attributed to additive quantities, but it is not correct. The value of an additive quantity for a whole object is equal to the sum of its values for the parts of the object in any division of the object into parts. The entropy of an ideal gas in classical thermodynamics is expressed by the equation that contains the term $Rn \ln(V/n)$, where $n$ is the number of moles of gas, $V$ is the volume of gas, $R$ is the universal gas constant, or by equations equivalent to it. As a result, the entropy of an ideal gas is equal to the sum of the entropies of its parts only if the parts of the gas are in different places (separated by an impermeable partition). If the parts of the gas form a mixture, then the sum of the entropies of the parts is not equal to the entropy of the gas. Despite this, the entropy of an ideal gas is considered to be an additive quantity. This gives rise to a series of inexplicable conclusions known as various formulations of the Gibbs paradox.

## Contents

# 1. Introduction

There are several formulations of the Gibbs paradox that appear in reasoning based on the principles of classical thermodynamics. The most well-known formulation we are talking to the paradoxical behavior of the entropy of the mixing of ideal gases during the transition from mixing different gases to mixing identical gases (see, e. g., [1–16]). Let us assume, that two different, chemically non-interacting ideal gases with equal temperature and pressure are separated by an impermeable partition. After the partition is removed, the gases mix, and the entropy of the system increases by the value of the entropy of mixing, which, according to calculations, does not depend on the nature and degree of difference of gases. If identical gases with the same temperatures and pressures, initially separated by a partition, mix, then the entropy of mixing is zero. Thus, supposing that the properties of mixing gases approach each other, then the entropy of mixing remains constant as long as there is some, even an infinitesimal, difference between the gases and turns to zero by a jump when the gases become identical. Since the entropy of mixing changes to zero from a value that does not depend on the properties of gases, then the magnitude of the jump in the entropy of mixing does not depend on how and how much the gases initially differed.

This behavior of the entropy of mixing is paradoxical. In classical thermodynamics, jumps in thermodynamic functions are always associated with jumps in the parameters of thermodynamic systems. In the above reasoning, it was assumed that there was a continuous (without a jump) transition from different to identical gases. A jump in the entropy of mixing under this condition looks impossible. If we assume that the jump in the entropy of mixing is associated with a jump of some parameter, then it is not known with the jump of which parameter. In addition, it looks mysterious that the magnitude of the jump in the entropy of mixing does not depend on the nature and degree of difference between gases.

Another formulation of this paradox concerns a similar behavior of the entropy of a mixture of ideal gases: when transitioning from a mixture of different gases to a mixture of identical gases by converging property values of gases, the value of the entropy of the system changes abruptly at the moment of transition from different to identical gases [17].

There are also other formulations of the Gibbs paradox that appear in the framework of classical thermodynamics (see, e.g., [6,15,18–24]). They will be discussed further. There are also formulations that appear and are discussed within the framework of statistical physics (see, e.g., [6,8,13–16]). They are not considered in this article.

The Gibbs paradox has been known for over a hundred years. It was discussed by many well-known physicists, including J. W. Gibbs, M. Planck, J. D. van der Waals, H. A. Lorentz, and P. W. Bridgman (see [1,2,8,12,18,22,24]). Its explanations have long been stated in textbooks (see, e.g., [2–5,7]); as a result, many physicists believe that this paradox has long been resolved. However, this is not the case. Different textbooks present different explanations, and new articles that discuss the paradox appear repeatedly (see, e.g., [9–16]). It can be stated that there is still no generally accepted explanation of this paradox.

This situation developed largely because the authors who discussed the Gibbs paradox did not pay due attention to its mathematical side. They did not take into account the fact that quantities whose behavior seems paradoxical can be expressed by equations. They discussed the behavior of the functions of many variables without taking into account the properties of the equations by which these functions are expressed. As a result, they did not notice that the appearance of various formulations of the Gibbs paradox in the framework of classical thermodynamics was due to the non-additivity of the entropy of an ideal gas. Since the entropy of an ideal gas is erroneously attributed to additive quantities, these formulations remain inexplicable. To demonstrate this, let us investigate the additivity of the entropy of an ideal gas and analyze the reasoning in which various formulations of the Gibbs paradox arise in the framework of classical thermodynamics.

## 2. On additive quantities, functions, properties

The quantity is called additive if its value $A$, corresponding to a whole object, is equal to the sum of its values $A_i$, corresponding to the parts of the object for any division of the object into parts (see, e.g., [25]):

$$A = \sum A_i . \tag{2.1}$$

Examples of additive quantities are the length of a line, surface area, and volume of a solid.

Additive quantities and additive properties of substances are mass, weight, number of moles, number of particles, heat capacity and some other properties of substances. For these properties of a certain amount of substance, for any division of this amount of substance into parts, the relations, that are particular cases of Eq. (2.1), holds true.

A function is called an additive function if it satisfies the additive Cauchy functional equation (see, e.g., [26,27]):

$$f(y_1 + y_2) = f(y_1) + f(y_2) . \tag{2.2}$$

If $y_1$ and $y_2$ are real numbers and $f(y)$ is a continuous function, then the solution of the additive Cauchy functional equation is a linear function [26,27]:

$$f(y) = by. \tag{2.3}$$

The additive properties of pure substances are the additive functions of the number of moles $n$ of substances since for any of these properties, the following relation holds true:

$$A(n_1 + n_2) = A(n_1) + A(n_2), \tag{2.4}$$

where $A(n_1 + n_2)$ is the value of the property $A$, corresponding to the $n_1 + n_2$ moles of pure substance, and $A(n_1)$ and $A(n_2)$ are the values of the property $A$, corresponding to $n_1$ and $n_2$ moles of substance.

For the additive properties of substances, the equation that is a special case of Eq. (2.3) is valid:

$$A = a_i n_i, \tag{2.5}$$

where $a_i$ is a molar value of the additive property $A$ for the pure $i$th substance.

$$a_i = \frac{A}{n_i}. \tag{2.6}$$

In ideal mixtures, the molar and specific properties of substances have the same values as in pure substances. The value of the additive property of an ideal mixture of substances $A_{mix}$ is equal to the sum of the additive properties of the substances that form the mixture. It follows from (2.1) and (2.5) for an ideal mixture of substances 1 and 2:

$$A_{mix} = a_1 n_1 + a_2 n_2 = a_{mix}(n_1 + n_2) = a_{mix} n_{mix}, \tag{2.7}$$

where $a_{mix}$ is an average molar value of the additive property $A$ for a mixture and $n_{mix}$ is the total number of moles of substances in the mixture.

$$n_{mix} = n_1 + n_2, \tag{2.8}$$

$$a_{mix} = \frac{A_{mix}}{n_{mix}} = \frac{a_1 n_1 + a_2 n_2}{n_1 + n_2} = a_1 x_1 + a_2 x_2, \tag{2.9}$$

$$x_i = \frac{n_i}{n_1 + n_2}. \tag{2.10}$$

A special case of Eq. (2.9) is the equation for the molar heat capacity at a constant volume of a mixture of ideal gases 1 and 2, $c_{Vmix}$:

$$c_{Vmix} = \frac{n_1 c_{V1} + n_2 c_{V2}}{n_1 + n_2} = \frac{n_1 c_{V1} + n_2 c_{V2}}{n_{mix}}, \tag{2.11}$$

where $c_{V1}$ and $c_{V2}$ are the molar heat capacities at a constant volume of gases 1 and 2.

According to (2.7), the value of the additive property of an ideal mixture $A_{mix}$ is a linear function of the number of moles of substances in the mixture $n_i$. It follows from (2.7) that the value of additive property $A$ of a mixture of $n_1$ and $n_2$ moles of identical substances (*i*th and *i*th) is equal to the value of property $A$ of $n_1 + n_2$ moles of pure *i*th substance.

We emphasize, that the additivity of properties, i.e., the fulfillment of relations (2.1) and (2.7), is only true for ideal mixtures of substances. In non-ideal mixtures, deviations from additivity are observed: the value of the sum of the additive properties of substances forming a non-ideal mixture differs from the value of the corresponding property of the mixture found by measurements. Deviations from additivity are explained within the framework of molecular theory by the fact that there are interactions in non-ideal mixtures between particles of different substances that do not exist between particles of the same substance and do not exist between particles of different substances in ideal mixtures.

### 3. On the additivity of the entropy of an ideal gas

The additivity of the mass, weight, heat capacity, and a number of other properties of substances was established and can be experimentally tested — according to measurements of the values of the corresponding values for a certain amount of a substance and for parts of this amount of a substance.

The additivity of the entropy of substances cannot be verified experimentally since entropy is a physical quantity that cannot be measured directly. However, entropy is a function of state, i.e., a certain function of the parameters of thermodynamic systems. In thermodynamics, an equation that expresses the entropy of an ideal gas in terms of its parameters is known. Using this equation, one can compare the entropy of a certain amount of an ideal gas and the sum of the entropies of the parts of this amount without making measurements.

The entropy of ideal gases in thermodynamics courses and in papers dealing with the Gibbs paradox (see, e.g., [1–6,10,17–21]) is expressed by the equation of the form (3.1) or equivalent to it:

$$S_i(n_i, V_i, T) = n_i \left( c_{Vi} \ln T + R \ln \frac{V_i}{n_i} + s_{0i} \right), \tag{3.1}$$

where $S_i(n_i, V_i, T)$ is the entropy of $n_i$ moles of the *i*th ideal gas, $T$ is the thermodynamic (absolute) temperature of the gas, $V_i$ is the volume of the *i*th gas, $c_{Vi}$ is the molar heat capacity of the $i$th ideal gas at a constant volume, $R$ is the universal gas constant, and $s_{0i}$ is the constant that depends on the nature of the gas.

In Eq. (3.1) and further, the entropy of some amount of the $i$th gas is denoted as the function of the quantities that unequivocally define its value, e.g., $S_i(n_1,V_1,T)$, $S_i(n_1+n_2,V_1+V_2,T)$, $S_1(n_1,V_1,T)$, $S_2(n_2,V_2,T)$, $S_1(n_1,V_1+V_2,T)$.

Using Eq. (3.1), we calculate the sums of the entropies of two parts of the $i$th ideal gas for two methods of partitioning the gas into parts.

In the first partitioning method, the gas is divided into parts by a partition. One part of the gas contains $n_1$ moles of gas, and the other part contains $n_2$ moles of gas; $n_1+n_2=n_i$. The pressure and temperature of the parts of the gas are equal to the pressure and temperature of the gas ($p_1=p_2=p_i$; $T_1=T_2=T$). The volumes of the parts are equal to $V_1$ and $V_2$, and the volume of the gas is equal to the sum of the volumes of the parts: $V_1+V_2=V_i$.

In this case, the entropies of the parts of the gas are equal to $S_i(n_1,V_1,T)$ and $S_i(n_2,V_2,T)$, and their sum is expressed by the equation that follows from (3.1):

$$S_i(n_1,V_1,T)+S_i(n_2,V_2,T)=(n_1+n_2)c_{Vi}\ln T+n_1 R\ln\frac{V_1}{n_1}+n_2 R\ln\frac{V_2}{n_2}+(n_1+n_2)s_{0i} \qquad (3.2)$$

Under conditions $p_1=p_2$ and $T_1=T_2$ from the equation of the state of an ideal gas $p_i V_i=n_i RT$, it follows:

$$\frac{V_1}{n_1}=\frac{V_2}{n_2}=\frac{V_1+V_2}{n_1+n_2}. \qquad (3.3)$$

It follows from (3.2) and (3.3):

$$S_i(n_1,V_1,T)+S_i(n_2,V_2,T)=n_i\left(c_{Vi}\ln T+R\ln\frac{V_1+V_2}{n_1+n_2}+s_{0i}\right)=S_i(n_1+n_2,V_1+V_2,T) \qquad (3.4)$$

Eq. (3.4) shows that when a gas is divided into parts by an impermeable partition, the sum of the entropies of the parts of the gas ($n_1$ and $n_2$ moles) is equal to the entropy of the $n_1+n_2$ moles of the gas. It should also be noted that under the conditions $p_1=p_2$ and $T_1=T_2$ from Eqs. (3.1) and (3.3) it follows that $S_i(n_i,V_i,T_i)$ is a linear function of the additive quantity $n_i$; therefore, it is an additive function and an additive quantity.

In the second method of partitioning, two parts of the $i$th gas ($n_1$ and $n_2$ moles) form a mixture, the volume of which is equal to $V_1 + V_2$. The volumes of the parts are equal to the volume of the mixture, and the temperature of the parts of the gas is equal to the temperature of the gas $T$.

In this case, the entropies of the parts of the $i$th gas are equal to $S_i(n_1, V_1+V_2, T)$ and $S_i(n_2, V_1+V_2, T)$, and their sum is expressed by the equation that follows from Eq. (3.1):

$$S_i(n_1, V_1+V_2, T) + S_i(n_2, V_1+V_2, T) =$$
$$= (n_1+n_2)c_{Vi} \ln T + n_1 R \ln \frac{V_1+V_2}{n_1} + n_2 R \ln \frac{V_1+V_2}{n_2} + (n_1+n_2)s_{0i}. \quad (3.5)$$

We transform Eq. (3.5) so that it contains a term equal to the entropy of $n_1 + n_2$ moles of the $i$th gas with volume $V_1 + V_2$:

$$S_i(n_1, V_1+V_2, T) + S_i(n_2, V_1+V_2, T) =$$
$$= (n_1+n_2)\left(c_{Vi} \ln T + R \ln \frac{V_1+V_2}{n_1+n_2} + s_{0i}\right) + R(n_1+n_2)\ln(n_1+n_2) - Rn_1 \ln n_1 - Rn_2 \ln n_2 = \quad (3.6)$$
$$= S_i(n_1+n_2, V_1+V_2, T) + L_x(n_1, n_2)$$

where

$$L_x(n_1, n_2) = R[(n_1+n_2)\ln(n_1+n_2) - (n_1 \ln n_1 + n_2 \ln n_2)] =$$
$$= -R(n_1+n_2)(x_1 \ln x_1 + x_2 \ln x_2) \quad (3.7)$$

where $x_1$ and $x_2$ are determined by Eq. (2.10).

If $n_1 = n_2 = 1$, then $L_x(n_1, n_2) = 2R\ln 2$.

The term $L_x(n_1, n_2)$ is called the logarithmic or concentration term [17,18].

As seen from Eq. (3.6), the appearance of the term $L_x(n_1, n_2)$ in the equation for the sum of the entropies of the parts of the gas that form the mixture is because $n_1 \ln n_1 + n_2 \ln n_2 \neq (n_1+n_2)\ln(n_1+n_2)$.

Equation (3.6) demonstrates that in the case when the parts of an ideal gas form a mixture, the sum of the entropies of the parts of the gas is not equal to the entropy of an ideal gas. Therefore, the entropy of an ideal gas is not an additive quantity and is not an additive property of an ideal gas.

At the same time, the entropy of an ideal gas is an extensive property of an ideal gas. Extensive properties of substances are those that are directly proportional to the amount of substance under conditions of an unchanged state of a thermodynamic system, in the case of homogeneous systems – of unchanged values of temperature and pressure. It follows from Eq. (3.1) that under this condition, the entropy of an ideal gas is directly proportional to its quantity $n_i$; therefore, it is an extensive property of an ideal gas.

We draw attention to the difference between the non-additivity of the entropy of an ideal gas in the case when its parts form a mixture and the non-additivity of the properties of substances in the case of non-ideal mixtures. Deviations from additivity in non-ideal mixtures are because, in non-ideal mixtures between particles of substances, there are such interactions that are not in ideal mixtures. Of course, the parts of an ideal gas form an ideal mixture. The non-additivity of the entropy of an ideal gas is because equation (3.1) contains the term $Rn_i \ln(V_i / n_i)$, which is not an additive quantity.

It should be noted that Eq. (3.1) contains two terms, $n_i c_{Vi} \ln T$ and $n_i s_{0i}$, that are linear functions of $n_i$, and they are additive functions. These terms are additive quantities: as seen from (3.2) and (3.5), the sums of the terms $n_i c_{Vi} \ln T$ and $n_i s_{0i}$ do not depend on the method by which the gas is divided into parts.

That the entropy of an ideal gas is equal to the sum of the entropies of its parts with one method of gas partitioning and is not equal to the sum of the entropies of its parts with another method of gas partitioning is not a unique property of entropy. The volume of a gas has the same property. The volume of a gas is equal to the sum of the volumes of the parts of the gas if the parts of the gas are separated by an impermeable partition and is not equal to the sum of the volumes of the parts if the parts of the gas form a mixture.

The pressure of an ideal gas is also equal to the sum of the pressures of its parts, not in every division of the gas into parts. The pressure of an ideal gas is equal to the sum of the pressures of its parts in the case when the parts of the gas form a mixture and is not equal to the sum of the pressures of the parts if the parts of the gas are separated by an impermeable partition.

Let us make two more remarks.

A mixture of $n_1$ and $n_2$ moles of the $i$th ideal gas is a mixture of $n_1$ and $n_2$ moles of identical gases ($i$th and $i$th). Thus, Eq. (3.6) expresses the sum of the entropies of $n_1$ and $n_2$ moles of identical gases ($i$th and $i$th) forming a mixture. Eq. (3.6) shows that the quantity "the sum of entropies of $n_1$ and $n_2$ moles of identical gases forming a mixture" $S_i(n_1, V_1 + V_2, T) + S_i(n_2, V_1 + V_2, T)$ and the quantity "entropy of $n_1 + n_2$ moles of an ideal gas" $S_i(n_1+n_2,V_1+V_2,T)$ are different functions of the same variables and parameters ($n_1, n_2, V_1, V_2, T, c_{Vi}, s_{0i}$) since the first equals the sum of the second one and the function $L_x(n_1,n_2)$.

If we compare Eqs. (3.4) and (3.6), we can conclude that the quantity "the sum of the entropies of identical gases forming a mixture" $S_i(n_1,V_1+V_2,T)+S_i(n_2,V_1+V_2,T)$ and the quantity "the sum of the entropies of identical ideal gases separated by a partition" $S_i(n_1,V_1,T)+S_i(n_2,V_2,T)$ are different functions of the same variables and parameters ($n_1, n_2, V_1, V_2, T, c_{Vi}, s_{0i}$) since the first is equal to the sum of the second and the function $L_x(n_1,n_2)$. Therefore, the difference between the values of quantities $S_i(n_1,V_1+V_2,T)+S_i(n_2,V_1+V_2,T)$ and $S_i(n_1,V_1,T)+S_i(n_2,V_2,T)$ cannot be interpreted as a change in the value of some function.

## 4. On the sum of the entropies of different ideal gases

Using Eq. (3.1), let us calculate the values of the sum of the entropies of two different ideal gases (1 and 2) for the case when the gases form a mixture and for the case when they are separated by an impermeable partition.

For the case when $n_1$ moles of gas 1 and $n_2$ moles of gas 2 form a mixture, the volume of which is equal to $V_1+V_2$, and the temperature is $T$, from (3.1), the equation for the sum of the entropies of these gases follows:

$$S_1(n_1,V_1+V_2,T)+S_2(n_2,V_1+V_2,T)=$$
$$=(n_1 c_{V1}+n_2 c_{V2})\ln T+n_1 R\ln\frac{V_1+V_2}{n_1}+n_2 R\ln\frac{V_1+V_2}{n_2}+n_1 s_{01}+n_2 s_{02}. \tag{4.1}$$

We transform Eq. (4.1):

$$S_1(n_1,V,T)+S_2(n_2,V,T)=$$
$$=(n_1+n_2)\left(c_{Vmix}\ln T+R\ln\frac{V_1+V_2}{n_1+n_2}+s_{0mix}\right)+R(n_1+n_2)\ln(n_1+n_2)-Rn_1\ln n_1-Rn_2\ln n_2= \tag{4.2}$$
$$=(n_1+n_2)\left(c_{Vmix}\ln T+R\ln\frac{V_1+V_2}{n_1+n_2}+s_{0mix}\right)+L_x(n_1,n_2)$$

where $c_{Vmix}$ is expressed by equation (2.11), $L_x(n_1,n_2)$ is expressed by equation (3.7), and $s_{0mix}$ is expressed by Eq. (4.3):

$$s_{0mix}=\frac{n_1 s_{01}+n_2 s_{02}}{n_1+n_2} \tag{4.3}$$

For the case, when $n_1$ moles of gas 1 and $n_2$ moles of gas 2 with the same pressures and temperatures and volumes $V_1$ and $V_2$ are separated by an impermeable partition, from (3.1), the equation for the sum of the entropies of these gases follows:

$$S_1(n_1,V_1,T)+S_2(n_2,V_2,T)=(n_1 c_{V1}+n_2 c_{V2})\ln T+n_1 R\ln\frac{V_1}{n_1}+n_2 R\ln\frac{V_2}{n_2}+n_1 s_{01}+n_2 s_{02}. \qquad (4.4)$$

It follows from (4.4) and (3.3), taking into account (2.11) and (4.3):

$$S_1(n_1,V_1,T)+S_2(n_2,V_2,T)=(n_1+n_2)\left(c_{V\text{mix}}\ln T+R\ln\frac{V_1+V_2}{n_1+n_2}+s_{0\text{mix}}\right). \qquad (4.5)$$

Eqs. (4.2) and (4.5) show that the sum of the entropies of different ideal gases forming a mixture $S_1(n_1,V_1+V_2,T)+S_2(n_2,V_1+V_2,T)$ exceeds the sum of the entropies of the same gases separated by a partition $S_1(n_1,V_1+V_2,T)+S_2(n_2,V_1+V_2,T)$ by a value $L_x(n_1,n_2)$. This demonstrates once again that the entropy of an ideal gas is not an additive quantity.

We draw attention to the fact that the appearance of term $L_x(n_1,n_2)$ in Eq. (4.2) for the sum of the entropies of different ideal gases is due to the same reason that the appearance of term $L_x(n_1,n_2)$ in Eq. (3.6) for the sum of the entropies of identical ideal gases. This reason is the non-additivity of the term $Rn_i\ln(V_i/n_i)$ in Eq. (3.1) under the condition that the volumes of the parts are equal to the volume of the gas.

If we compare Eqs. (4.1) and (4.4), we can conclude that the quantity "the sum of the entropies of different ideal gases forming a mixture" $S_1(n_1,V_1+V_2,T)+S_2(n_2,V_1+V_2,T)$ and the quantity "the sum of the entropies of different ideal gases separated by a partition" $S_1(n_1,V_1,T)+S_2(n_2,V_2,T)$ are different functions of the same variables and parameters ($n_1$, $n_2$, $V_1$, $V_2$, $T$, $c_{V1}$, $c_{V2}$, $s_{01}$, $s_{02}$). The same conclusion follows from Eqs. (4.2) and (4.5), according to which the function $S_1(n_1,V_1+V_2,T)+S_2(n_2,V_1+V_2,T)$ is equal to the sum of the function $S_1(n_1,V_1,T)+S_2(n_2,V_2,T)$ and the function $L_x(n_1,n_2)$. Therefore, the difference between the values of quantities $S_1(n_1,V_1+V_2,T)+S_2(n_2,V_1+V_2,T)$ and $S_1(n_1,V_1,T)+S_2(n_2,V_2,T)$ cannot be interpreted as a change in the value of some function.

At the same time, the sum of the entropies of different gases separated by a partition $S_1(n_1,V_1,T)+S_2(n_2,V_2,T)$ is a function of the same type as the sum of the entropies of identical gases separated by a partition $S_i(n_1,V_1,T)+S_i(n_2,V_2,T)$. These functions differ only in parameter values: Eq. (3.4) is a particular case of Eq. (4.5) under the condition $c_{V1} = c_{V2} = c_i$, $s_{01} = s_{02} = s_{0i}$.

The sum of the entropies of different ideal gases forming a mixture $S_1(n_1, V_1+V_2, T) + S_2(n_2, V_1+V_2, T)$ is a function of the same type as the sum of the entropies of identical gases forming a mixture $S_i(n_1, V_1+V_2, T) + S_i(n_2, V_1+V_2, T)$ and at the same time a function of a different type than the entropy of an ideal gas $S_i(n_1+n_2, V_1+V_2, T)$. A special case of Eq. (4.4) under condition $c_{V1} = c_{V2} = c_i$, $s_{01} = s_{02} = s_{0i}$ is Eq. (3.6) but not an equation of the form (3.1).

## 5. Detecting errors in reasoning leading to various formulations of the Gibbs paradox in classical thermodynamics

So, the entropy of an ideal gas, which is expressed by equation (3.1) containing the term $Rn_i \ln(V_i/n_i)$, is not an additive quantity. However, it is erroneously attributed to additive quantities, and it is assumed that the entropy of a thermodynamic system consisting of several different ideal gases is equal to the sum of the entropies of these gases in the case when the gases are separated by impermeable partitions (see, e.g., [1–6,10,17–21]) and in the case when they form a mixture (see, e.g., [1–6,10,17–21]). The second statement is known as the Gibbs theorem on the entropy of a mixture of ideal gases (see, e.g., [2–4,6,17]).

The first statement for a system consisting of two ideal gases separated by an impermeable partition can be expressed by the equation:

$$S_{part}(n_1, n_2, V_1, V_2, T) = S_1(n_1, V_1, T) + S_2(n_2, V_2, T), \qquad (5.1)$$

where $S_{part}(n_1, n_2, V_1, V_2, T)$ is the entropy of the system consisting of $n_1$ moles of gas 1 and $n_2$ moles of gas 2 separated by an impermeable partition, the volumes of which are equal to $V_1$ and $V_2$, and the temperature is $T$.

The second statement for a system that is a mixture of two ideal gases can be expressed by the equation:

$$S_{mix}(n_1, n_2, V_1+V_2, T) = S_1(n_1, V_1+V_2, T) + S_2(n_2, V_1+V_2, T), \qquad (5.2)$$

where $S_{mix}(n_1, n_2, V_1+V_2, T)$ is the entropy of the mixture of $n_1$ moles of gas 1 and $n_2$ moles of gas 2, the volume of which is $V_1+V_2$, and the temperature is $T$.

If two ideal gases separated by a partition have the same temperature and pressure, then it follows from Eqs. (5.1) and (4.5):

$$S_{part}(n_1, n_2, V_1, V_2, T) = (n_1 + n_2)\left(c_{Vmix} \ln T + R \ln \frac{V_1 + V_2}{n_1 + n_2} + s_{0mix}\right). \quad (5.3)$$

It follows from Eqs. (5.2) and (4.2) for a mixture of two ideal gases:

$$S_{mix}(n_1, n_2, V_1 + V_2, T) = (n_1 + n_2)\left(c_{Vmix} \ln T + R \ln \frac{V_1 + V_2}{n_1 + n_2} + s_{0mix}\right) + L_x(n_1, n_2). \quad (5.4)$$

Ideal gases 1 and 2, whose separated by an impermeable partition, whose pressures are $p$, temperatures are $T$, and volumes are $V_1$ and $V_2$, and the same gases forming a mixture whose pressure is $p$, temperature is $T$, and volume is $V_1 + V_2$ — these are two ways of dividing into parts of the system, consisting of two ideal gases 1 and 2, the volume of which is $V_1 + V_2$, temperature is $T$, pressure is $p$. According to (5.3) and (5.4), $S_{mix}(n_1, n_2, V_1 + V_2, T) \neq S_{part}(n_1, n_2, V_1, V_2, T)$. It follows from this that the entropy of an ideal gas is not an additive quantity since the sum of the values of the additive quantity corresponding to the parts of the object does not depend on the way the object is divided into parts. However, the entropy of an ideal gas is erroneously attributed to additive quantities. As a result, in a number of reasoning within the framework of classical thermodynamics, in which the entropy of pure ideal gas and the sum of the entropies of ideal gases forming a mixture are used, inexplicable conclusions arise that make up the content of various formulations of the Gibbs paradox.

Let us briefly present the essence of these formulations using the equations obtained above.

It is considered paradoxical in one formulation that the equation of the form (5.4) for the entropy of a mixture of different ideal gases does not pass into the equation of the form (3.1) for the entropy of a pure ideal gas in the case when the mixture consists of identical gases. In another formulation, it is considered paradoxical that the sum of entropies of identical ideal gases forming a mixture, which is expressed an equation of the form (5.4), exceeds by the quantity $L_x(n_1, n_2)$ the sum of entropies of the same gases separated by a partition, which is expressed by an equation of the form (5.3). There are formulations in which it is considered paradoxical that the entropy of the system or the change in the entropy of the system decreases by $L_x(n_1, n_2)$ in the case when calculating these quantities, instead of Eq. (5.4), which contains the term $L_x(n_1, n_2)$, equation of the form (3.1) is used, in which there is no term $L_x(n_1, n_2)$. There are other formulations. However, in all formulations of the Gibbs paradox that arise in the framework of classical thermodynamics, the entropy of a mixture of ideal gases and the quantity $L_x(n_1,n_2)$ appear. Since no one associates the appearance of the term $L_x(n_1,n_2)$ in various equations with the non-additivity of the entropy of an ideal gas, these various formulations of the Gibbs paradox remain unexplained.

Let us analyze the reasoning in which, within the framework of classical thermodynamics, various formulations of the Gibbs paradox appear, and we will reveal the errors in these reasoning.

J. D. van der Waals and Ph. Konstamm [18] obtained an equation for the entropy of the $i$th ideal gas, which contains a term equivalent to the $Rn_i \ln(V_i/n_i)$ term. Assuming that "in very rarefied gases, all properties are additive", they found the entropy of a mixture of two ideal gases by adding the entropies of these gases and obtained an equation that contains a logarithmic term equivalent to $L_x(n_1,n_2)$. Then, in the "The Gibbs Paradox" section, they found the sum of the entropies of the identical gases that form the mixture and obtained a result similar to that expressed by Eq. (3.6): the entropy of a mixture of identical ideal gases is greater than the entropy of a pure gas by the value of the logarithmic term. J. D. van der Waals and Ph. Konstamm called this result paradoxical. This result is caused by the non-additivity of the entropy of an ideal gas and seems paradoxical (inexplicable) to someone who erroneously believes that the entropy of an ideal gas is an additive quantity.

A similar formulation of the Gibbs paradox can be found in the book by M. A. Leontovich [23]: the sum of free energies of $n_1$ and $n_2$ moles of identical ideal gases forming a mixture is not equal to the free energy of $n_1+n_2$ moles of pure gas.

This result is explained by the fact that the free energy of an ideal gas is not an additive quantity because its equation contains the term $-TS$, and the entropy of an ideal gas is not an additive quantity. Since Eq. (3.6) for the sum of entropies of identical ideal gases forming a mixture contains the term $L_x(n_1,n_2)$, then the sum of free energies of $n_1$ and $n_2$ moles of the $i$th ideal gas forming a mixture differs from the free energy of $n_1+n_2$ moles of the $i$th ideal gas by $-TL_x(n_1,n_2)$. This result seems paradoxical to those who erroneously consider the entropy and free energy of an ideal gas to be additive quantities.

Note that if the entropy and free energy of an ideal gas were additive quantities, then these two formulations of the Gibbs paradox would not arise. As mentioned above, from equation (2.7), which expresses the value of the additive property of a mixture of substances in terms of molar properties and quantities of substances, it follows that the value of the additive property of a mixture of $n_1$ and $n_2$ moles of identical substances (*i*th and *i*th) is equal to the value of the same property $n_1 + n_2$ moles of pure *i*th substance.

B. M. Kedrov discussed the formulation of the Gibbs paradox, in which refers to the paradoxical behavior of the entropy of a thermodynamic system during the transition from a mixture of different ideal gases to a pure gas by converging property values of gases under the condition of constancy of their quantities [17].

This formulation arises in such reasoning.

Suppose there is a mixture $n_1$ of moles of gas 1 and $n_2$ moles of gas 2. The entropy of gases is expressed by Eq. (3.1); the entropy of this thermodynamic system is equal to the entropy of the mixture, respectively, is equal to the sum of the entropies of gases 1 and 2 and is expressed by Eq. (4.2), which contains the term $L_x(n_1, n_2)$. If at constant values $n_1$ and $n_2$ the properties of gases converge, approaching the properties of the *i*th ideal gas, but the gases remain different, then the value of the entropy of the mixture, i.e., the sum $S_1(n_1, V_1 + V_2, T) + S_2(n_2, V_1 + V_2, T)$, is expressed by Eq. (4.2) and approaches the value of the sum of the entropies of identical gases $S_i(n_1, V_1 + V_2, T) + S_i(n_2, V_1 + V_2, T)$, which is expressed by Eq. (3.6). If gases 1 and 2 are somewhat different, but $c_{V1} = c_{V2} = c_{Vi}$, $s_{01} = s_{02} = s_{0i}$, then the entropy of the mixture is expressed by equation (3.6), which contains the term $L_x(n_1, n_2)$. When gases 1 and 2 become identical to the *i*th gas, the mixture turns into a pure *i*th gas, the entropy of which is expressed by Eq. (3.1) for the case when $n_i = n_1 + n_2$ and $V_i = V_1 + V_2$. Equation (3.1) does not contain the term $L_x(n_1, n_2)$, on the basis of which it is concluded that the entropy of the system decreases abruptly by $L_x(n_1, n_2)$.

Thus, the conclusion about the jump in the entropy of the system in this reasoning is because the entropy of a mixture of different ideal gases is expressed by Eq. (5.4), which contains the term $L_x(n_1, n_2)$, and the entropy of a mixture of identical ideal gases is expressed by Eq. (3.1) for the entropy of a pure ideal gas, which does not contain the term $L_x(n_1, n_2)$. This difference between Eqs. (4.2) and (3.1) is due to the non-additivity of the entropy of an ideal gas. It looks paradoxical (inexplicable) for those who erroneously attribute the entropy of an ideal gas to additive quantities.

A number of formulations of the Gibbs paradox arise when discussing the entropy of mixing of ideal gases separated by an initially impermeable partition.

Since entropy is a state function, the entropy of mixing can be expressed by the equation:

$$\Delta S = S_{II} - S_I, \tag{5.5}$$

where $S_I$ is the entropy of the system before mixing, $S_{II}$ is the entropy of the mixture.

This is how the entropy of mixing was found by many authors who discussed the Gibbs paradox (see, e.g., [1–6,10,17–21]).

In the case of mixing $n_1$ moles of gas 1 and $n_2$ moles of gas 2 with the same pressures and temperatures and volumes $V_1$ and $V_2$, separated by an initially impermeable partition, $S_I = S_{part}(n_1, n_2, V_1, V_2, T)$, $S_{II} = S_{mix}(n_1, n_2, V_1 + V_2, T)$.

For this case, it follows from Eqs. (5.1), (5.2), (5.5) that the entropy of mixing of different ideal gases $\Delta S_{mix}^d$ is expressed by the equation:

$$\Delta S_{mix}^d = [S_1(n_1, V_1 + V_2, T) + S_2(n_2, V_1 + V_2, T)] - [S_1(n_1, V_1, T) + S_2(n_2, V_2, T)]. \tag{5.6}$$

It follows from Eqs. (5.6), (4.2), (4.5):

$$\Delta S_{mix}^d = L_x(n_1, n_2). \tag{5.7}$$

If $n_1 = n_2 = 1$, then it follows from (5.7) and (3.7):

$$\Delta S_{mix}^d = 2R \ln 2. \tag{5.8}$$

After obtaining the results, which are expressed by Eqs. (5.7) or (5.8), some authors argue as follows. According to Eqs. (5.7) or (5.8), the entropy of mixing of different ideal gases with the same temperatures and pressures is equal to $L_x(n_1, n_2)$ or $2R \ln 2$ and does not depend on the properties of the mixed gases. Therefore, we can conclude that under the same conditions, the entropy of mixing of identical ideal gases is also equal to $L_x(n_1, n_2)$ or $2R \ln 2$. However, when identical gases are mixed, the final state of the system does not differ from the initial state. Therefore, the change in entropy when mixing identical gases is zero. A contradiction arises, the appearance of which these authors call the Gibbs paradox (see, e.g., [18–22]).

This formulation of the Gibbs paradox arises from an erroneous interpretation of Eqs. (5.6)–(5.8). In the above reasoning, it was assumed that Eqs. (5.6)–(5.8) express the entropy of mixing of ideal gases. However, it is not. Eqs. (5.6)–(5.8) demonstrate the non-additivity of the entropy of an ideal gas. The first term in square brackets in Eq. (5.6) is the sum of the entropies of the ideal gases forming a mixture. The second term in square brackets in Eq. (5.6) is the sum of the entropies of the same ideal gases separated by an impenetrable partition. If the entropy of an ideal gas were an additive quantity, these sums would have the same value, and their difference would be zero. After all, the sums of the additive properties of ideal gases (mass, number of particles, heat capacity, internal energy) have the same values for gases that form a mixture, and for the same gases separated by a partition.

However, the entropy of an ideal gas is not an additive quantity because Eq. (3.1) for the entropy of an ideal gas contains the term $Rn_i \ln(V_i/n_i)$. Using the term of K. G. Denbigh and M. L. G. Redhead [9], we can say that Eq. (5.6)–(5.8) express such a "mathematical artifact": if $\frac{V_1+V_2}{n_1+n_2} = \frac{V_1}{n_1} = \frac{V_2}{n_2}$, then

$$\left(n_1 R \ln \frac{V_1+V_2}{n_1} + n_2 R \ln \frac{V_1+V_2}{n_2}\right) - \left(n_1 R \ln \frac{V_1}{n_1} + n_2 R \ln \frac{V_2}{n_2}\right) = L_x(n_1, n_2).$$

Since the term $Rn_i \ln(V_i/n_i)$ does not depend on the properties of gases, the right-hand sides of Eqs. (5.6)–(5.8) have the same values for different and identical gases. This result appears paradoxical for those who erroneously attribute the entropy of an ideal gas to additive quantities and erroneously believe that Eqs. (5.6)–(5.8) express the entropy of mixing of ideal gases initially separated by a partition.

The formulation of the Gibbs paradox, similar to the previous one, was discussed by H. A. Lorentz. He wrote that when mixing one and one mole of different ideal gases initially separated by a partition, the free energy of the system decreases by an amount $2R \ln 2$ that does not depend on the type of gases. However, for identical gases, this conclusion is not applicable, since after the removal of the partition separating identical gases, the free energy does not change. Lorentz concluded that the case of identical gases cannot be considered the limiting case of different gases and called this conclusion paradoxical [24].

The appearance of a paradox here is because the free energy of an ideal gas is attributed to additive quantities, although it is not an additive quantity, since its equation contains a term $Rn_i T \ln(V_i/n_i)$. The sum of such terms is the same for different and identical gases. Therefore, there is nothing paradoxical in the fact that the difference between the sum of the free energies of ideal gases forming a mixture and the sum of the free energies of the same gases separated by a partition are the same for different and identical gases.

The conclusion that the change in free energy during mixing of ideal gases, initially separated by a partition, does not depend on their properties, is in sharp contradiction with the fact that the work required to separate a mixture of gases is the higher, the closer the properties of the separated gases are. Some authors have noted this contradiction and concluded that the conclusion that the change in free energy during the mixing of ideal gases does not depend on their properties is erroneous [11, 28]. However, they were unable to convincingly substantiate this conclusion because they did not pay enough attention to the mathematical aspects of obtaining this erroneous conclusion.

It should be mentioned that most authors assume that Eqs. (5.6)–(5.8) cannot be used to determine the entropy of mixing of identical gases separated before mixing by a partition, since when mixing identical gases, not a mixture is formed, but the pure gas. With this approach $S_I = S_i(n_1, V_1, T) + S_i(n_2, V_2, T)$ and $S_{II} = S_i(n_1 + n_2, V_1 + V_2, T)$ and for the entropy of mixing of identical ideal gases $\Delta S_{mix}^i$ from Eq. (5.5), it follows:

$$\Delta S_{mix}^i = [S_i(n_1 + n_2, V_1 + V_2, T)] - [S_i(n_1, V_1, T) + S_i(n_2, V_2, T)]. \tag{5.9}$$

It follows from Eqs. (5.9), (3.1), (3.4):

$$\Delta S_{mix}^i = 0. \tag{5.10}$$

According to Eq. (5.10), the entropy of mixing identical ideal gases with the same temperatures and pressures is zero. This result is consistent with the position that after the removal of the partition separating identical ideal gases with the same temperatures and pressures, the state of the thermodynamic system does not change. Accordingly, the problems that constitute the content of the two formulations of the Gibbs paradox discussed above do not arise.

However, in this case, a conclusion arises about a paradoxical jump in the entropy of the mixing of ideal gases in the transition from the mixing of different to the mixing of identical gases, which is the essence of the most well-known formulation of the Gibbs paradox [1–16]. This conclusion follows from a comparison of Eq. (5.7) or (5.8) with Eq. (5.10). The entropy of mixing of different ideal gases, according to Eqs. (5.7) and (5.8), does not depend on their properties and is equal to $L_x(n_1, n_2)$ or $2R \ln 2$. The entropy of mixing of identical ideal gases, according to Eq. (5.10), is equal to zero. In the transition from mixing of different to mixing of identical gases, the entropy of mixing decreases abruptly by $L_x(n_1, n_2)$ or $2R \ln 2$ and vanishes.

Assuming that the entropy of mixing cannot change abruptly if the parameters of the system change smoothly, various authors explained the jump in the entropy of mixing by various differences between the mixing of identical gases from the mixing of different gases or by various differences between a mixture of identical gases and a mixture of different gases (see [1–5,7,10,13,14]).

At the same time, they did not take into account the fact that the difference between the values of $\Delta S_{mix}^d$ and $\Delta S_{mix}^i$, equal to $L_x(n_1, n_2)$, is a function different from $\Delta S_{mix}^d$ and $\Delta S_{mix}^i$. It follows from this that the quantities $\Delta S_{mix}^d$ and $\Delta S_{mix}^i$ are different functions, since $\Delta S_{mix}^d$ is equal to the sum of two different functions: $\Delta S_{mix}^i$ and $L_x(n_1,n_2)$. Since $\Delta S_{mix}^d$ and $\Delta S_{mix}^i$ are different functions, it is impossible to carry out the transition from function $\Delta S_{mix}^d$ to function $\Delta S_{mix}^i$ by changing the values of the parameters on which depends $\Delta S_{mix}^d$. That is why all attempts to connect the paradoxical jump in the entropy of mixing with the jump in the gas difference parameter were unsuccessful.

If we compare Eqs. (5.6) and (5.9), which express $\Delta S_{mix}^d$ and $\Delta S_{mix}^i$, then we can see that the difference in the values of $\Delta S_{mix}^d$ and $\Delta S_{mix}^i$ equal to $L_x(n_1,n_2)$ is due to the difference in the first terms in square brackets of these equations. The first term in square brackets of Eq. (5.6) is the sum of the entropies $n_1$ and $n_2$ moles of two ideal gases that form the mixture whose equation contains the term $L_x(n_1,n_2)$. The first term in square brackets of Eq. (5.9) is the entropy $n_1+n_2$ of moles of a pure ideal gas, the equation of which does not contain the term $L_x(n_1,n_2)$. As shown above, this difference between the quantities $S_1(n_1,V_1+V_2,T)+S_2(n_2,V_1+V_2,T)$ and $S_i(n_1+n_2,V_1+V_2,T)$ is due to the non-additivity of the entropy of a pure ideal gas, the equation of which contains the term $Rn_i\ln(V/n_i)$. Accordingly, the difference in values $\Delta S_{mix}^d$ and $\Delta S_{mix}^i$ on the value $L_x(n_1,n_2)$ is also due to the non-additivity of the entropy of a pure ideal gas and is inexplicable for those who erroneously attribute the entropy of an ideal gas to additive quantities.

## 6. Conclusions

Various formulations of the Gibbs paradox in classical thermodynamics appear because the entropy of an ideal gas, which is expressed by an equation containing a term $Rn_i\ln(V_i/n_i)$ or equations equivalent to it, is erroneously attributed to additive quantities and is considered an additive property of ideal gases. Since the entropy of an ideal gas is not an additive quantity, in various arguments within the framework of classical thermodynamics, consequences appear that contradict the assumption of the additivity of the entropy of an ideal gas. These consequences turn out to be inexplicable for those who erroneously attribute the entropy of an ideal gas to additive quantities. If this error is not made, then the Gibbs paradox will not appear in classical thermodynamics.

### Acknowledgements
The author is grateful to professor Oleksandr Andriiko, senior lecturer Vasiliy Pikhorovich and the late associate professor Viktor Haidey for helpful discussions and comments.